\documentclass{article}
\usepackage{spconf,amsmath,graphicx}
\usepackage[hidelinks]{hyperref}

\title{AudioRWKV: Efficient and Stable Bidirectional RWKV for Audio Pattern Recognition}
\name{Jing Wang, Maoxiang Wu, Jiayu Xiong, Jianlong Kwan, Jun Xue
\thanks{Jing Wang, Maoxiang Wu, Jiayu Xiong and Jianlong Kwan is with the Department of Computer Science and Technology, Huaqiao University, Xiamen 361021, China (jiayu's e-mail: yuinst@outlook.com)} 
\thanks{Jun Xue is with the Wuhan University, Wuhan, China} 
\thanks{Code: \url{https://github.com/Jiayu-Xiong/AudioRWKV}.}
}
\address{Department of Computer Science and Technology, Huaqiao University and Wuhan University}
\usepackage{booktabs} 
\usepackage{amssymb}
\usepackage{multirow}
\usepackage{subfigure}

\begin{document}
%
\maketitle
\begin{abstract}
Recently, Transformers (e.g., Audio Spectrogram Transformers, AST) and state-space models (e.g., Audio Mamba, AuM) have achieved remarkable progress in audio modeling. However, the $\mathcal O(L^2)$ computational complexity of the Transformer architecture hinders efficient long-sequence processing, while the Mamba architecture tends to become unstable when scaling parameters and data.
To address these challenges, this paper proposes AudioRWKV (A-RWKV), a highly efficient and stable architecture for audio modeling. Specifically, we inherit the stable and efficient recurrent formulation of RWKV7 and replace its 1D token-shift operation with a 2D depthwise separable convolution to better capture local spectro-temporal patterns. Furthermore, we adapt the original causal WKV kernel into a bidirectional WKV kernel (Bi-WKV), enabling global context modeling over the entire audio sequence while maintaining linear computational complexity.
Benefiting from the inherent stability of the RWKV7 foundation, A-RWKV scales seamlessly to larger model sizes. Experimental results demonstrate that, under the same linear-model regime, A-RWKV-S (22M) achieves performance parity with AuM-B (92M) while exhibiting more stable throughput than AST; for long-form audio ($\sim$5 minutes 28 seconds), WKV7 achieves up to a 13.3$\times$ speedup in processing.
\end{abstract}

\section{Introduction}
\label{sec:intro}

Transformer-based architectures~\cite{vaswani2017attention, dosovitskiy2020image}, particularly the Audio Spectrogram Transformer (AST)~\cite{gong2021ast}, have established new performance benchmarks across a variety of audio understanding tasks, owing to their powerful global information processing capabilities. However, the self-attention mechanism, which underpins these models, has a computational and memory complexity that scales quadratically with the input sequence length ($\mathcal O(L^2)$)~\cite{sun2023retentive, gu2023mamba, peng2023rwkv}. This intrinsic limitation poses a significant barrier to their application in scenarios involving high-resolution or long-duration audio~\cite{li2024exploring, xiao2025rwkvtts}, making the exploration of more efficient architectures a critical area of research.

In recent developments, models with linear-time complexity, such as the state-space model Audio Mamba (AuM)~\cite{erol2024audio}, have emerged as compelling alternatives. These models demonstrate remarkable efficiency in processing long sequences and have achieved competitive results on several audio benchmarks. However, adapting these linear mechanisms for large-scale audio tasks is not straightforward. For instance, existing studies on AuM have primarily focused on small to medium-sized models. As attempts are made to scale them up to larger parameter counts and datasets, they tend to exhibit training instabilities, such as vanishing or exploding gradients, which hinders further performance improvement. Thus, effectively addressing the scalability and stability of these linear models remains a pivotal, unsolved challenge~\cite{sieber2024understanding, duan2024vision}.

The RWKV architecture~\cite{peng2023rwkv}, originating from the field of NLP, presents an attractive foundation for tackling these issues. It uniquely blends the linear complexity and constant memory usage of RNNs during inference with the parallelizable training of Transformers, has already been applied in the fields related to speech~\cite{li2024exploring, xiao2025rwkvtts} and vision~\cite{duan2024vision}. Within the RWKV family, the latest iteration, RWKV7~\cite{peng2025rwkv}, offers significant advancements. Its state transition matrix is more expressive than those of its predecessors, allowing for both exponential decay~\cite{sun2023retentive} and dynamic, per-channel updates. This sophisticated design enhances its modeling capacity while ensuring superior numerical stability during training, providing a robust starting point for adaptation to other domains.

Based on these observations, we propose AudioRWKV (A-RWKV). Our approach preserves the core efficiency and stability benefits of the RWKV7 architecture while incorporating essential modifications to process 2D audio spectrograms. We build upon the robust RWKV7 foundation, including its recurrent formulation and highly optimized CUDA kernel. Our novel contributions are primarily twofold: first, we replace the 1D token shift with a 2D depthwise separable convolution (DWConv2D) to effectively model local spectro-temporal patterns. Second, we adapt the original causal attention into a bidirectional global attention mechanism (Bi-WKV), allowing each time frame to attend to the entire audio context with linear complexity. By synergizing these audio-specific designs with the inherent stability of RWKV7, we successfully mitigate the scaling issues encountered by models like AuM.

In this paper, our main contributions are:
\begin{enumerate}
    \item We propose A-RWKV, a cost-effective and scalable backbone for audio tasks. It retains the global modeling strengths of AST while reducing computational complexity to a linear scale, offering an efficient solution for long-form audio processing.
    \item We develop a bidirectional global attention mechanism (Bi-WKV) combined with a DWConv2D-based token shift method. These operators, tailored for spectrograms, achieve effective and efficient feature aggregation across both local and global scopes.
    \item We demonstrate that by building on the stable RWKV7 foundation, A-RWKV overcomes the training instability issues that limit the scalability of competing linear-time models like AuM, enabling the successful training of larger models for superior performance.
\end{enumerate}

\section{Related Work}
The field of audio processing has been significantly advanced by Transformer-based models, most notably the Audio Spectrogram Transformer (AST)~\cite{gong2021ast}, which adapted the success of the Vision Transformer (ViT)~\cite{dosovitskiy2020image} to spectrograms. These models excel at capturing global context through self-attention but are inherently constrained by its quadratic complexity, hindering their application to long-form audio signals. This limitation has spurred research into efficient architectures with linear-time complexity. 

Among these, State-Space Models (SSMs) like Mamba~\cite{gu2023mamba} have gained prominence, leading to adaptations for audio such as Audio Mamba (AuM)~\cite{erol2024audio}. While computationally efficient, these models have demonstrated challenges in training stability when scaled to larger model sizes, a limitation also observed in their vision counterparts~\cite{zhu2024vision}. Concurrently, the RWKV architecture~\cite{peng2023rwkv} offers an alternative path, reformulating attention into a parallelizable recurrent neural network (RNN) that combines linear complexity with robust performance in natural language processing. Other approaches such as Retentive Networks (RetNet)~\cite{sun2023retentive} have also explored novel trade-offs between parallelizability, recurrence, and performance.

\begin{figure}[t]
    \centering
    \includegraphics[width=1\linewidth]{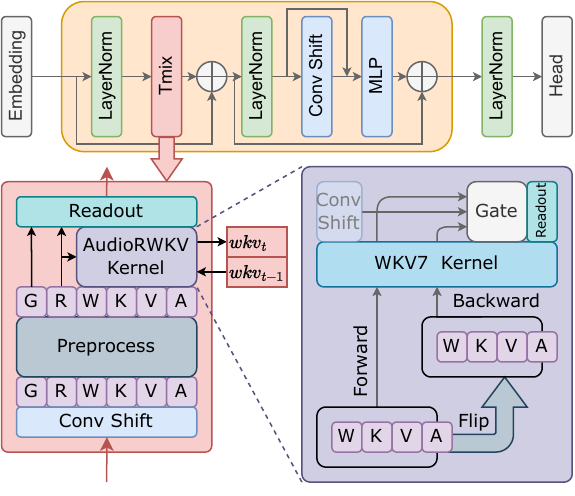}
    \caption{Overview of A-RWKV. The "Preprocess" and "Readout" structure is exactly the same as that of RWKV7. "Conv Shift" is 2D DWConv.}
    \label{fig:placeholder}
\end{figure}

\section{Methodology}

This work adapts the RWKV7 architecture~\cite{peng2025rwkv} for processing 2D spatial data, exemplified by audio Mel spectrograms. The foundational principle is the systematic replacement of the original 1D token shift operation with a 2D depthwise separable convolution (DWConv2D). This substitution enables the model to capture local spatial context, which is then integrated into the recurrent state dynamics through a bidirectional scanning mechanism.

The model first processes an input Mel spectrogram $S \in \mathbb{R}^{B \times 1 \times H \times W}$ using a convolutional embedding layer. This layer functions as a patch embedding operator, producing a sequence of patch features $x_{emb} \in \mathbb{R}^{B \times L \times D}$. A learnable absolute positional embedding $\mathbf{P}_{pos} \in \mathbb{R}^{1 \times L \times D}$ is added to this sequence to provide spatial awareness.
\begin{equation}
x_0 = \mathrm{Flatten}(\mathrm{Conv2D}(S)) + \mathbf{P}_{pos}
\end{equation}
The resulting sequence $x_0$ is then processed by a stack of $N$ identical blocks, each composed of a spatial mixing module and a channel mixing module, arranged with pre-LayerNorm residual connections.

The core of our adaptation lies within the Bidirectional Spatial Mixing module, which re-purposes the RWKV7 Time Mixing mechanism for spatial data. To capture local context, the input sequence $x$ is first reshaped into its two-dimensional form $X \in \mathbb{R}^{B \times D \times H_i \times W_i}$. A DWConv2D is applied, and the difference from the original input forms a local residual feature $x_{res}$.
\begin{equation}
x_{res} = \mathrm{Flatten}(\mathrm{DWConv2D}(X) - X)
\end{equation}
This local residual is then used to generate locally-aware inputs for all dynamic parameters via a channel-wise interpolation, governed by learnable vectors $\mathbf{\mu}_{\square}$.
\begin{equation}
x^{\square}_{t} = x_t + x_{res,t} \odot \mathbf{\mu}_{\square}
\end{equation}
Based on these inputs, the computation of all dynamic parameters such as the decay $w_t$, receptance $r_t$, key $k_t$, and value $v_t$, as well as the recursive evolution of the Weighted Key Value (WKV) state $\mathbf{wkv}_t$, strictly adhere to the formulations presented in the original RWKV7 model.
\begin{equation}
\mathbf{wkv}_t = \mathbf{wkv}_{t-1} f(w_t, \kappa_t, a_t) + g(v_t, \tilde{k}_t)
\end{equation}
This recurrence operates in parallel on both the forward sequence ($t=1, \dots, L$) and a time-reversed backward sequence, yielding two output sequences $p^{\rightarrow}$ and $p^{\leftarrow}$. A dynamic gate $\mathbf{G}$, derived from the local residual $x_{res}$, is used to fuse these two contextual representations.
\begin{equation}
p_{fused} = \mathbf{G} \odot p^{\rightarrow} + (1 - \mathbf{G}) \odot \mathrm{Flip}(p^{\leftarrow}, \mathrm{dim}=1)
\end{equation}
The Channel Mixing module, a feed-forward network, is similarly modified. Its internal token shift operation is also replaced by the same DWConv2D-based local residual mechanism.

After processing through all $N$ blocks, the final output sequence $x_N$ undergoes a final LayerNorm. Global average pooling is then applied across the sequence dimension to produce a single feature vector per sample. This vector is passed to a final linear layer to generate the classification logits
\begin{equation}
Y_{pred} = \mathrm{FC}(\mathrm{Mean}(\mathrm{LayerNorm}(x_N), \mathrm{dim}=1)),
\end{equation}
then calculate soft target cross-entropy loss and complete the backpropagation.

\section{Experiments}
In this section, we aim to demonstrate the advantages of Bi-WKV as sequence modeling operator over attention and Bi-SSM.

\subsection{Settings, Datasets and Baselines.}

\textbf{Model Settings.}
AudioRWKV(-B) is a 12-layer model with a 768d embedding and learnable absolute positional embeddings. All other architectural details and parameter initialization strictly follow the original RWKV-7 specification. We train all models for 25 epochs using the AdamW~\cite{adamw} optimizer with a base learning rate of 2e-5, a batch size of 1024, and a learning rate schedule of linear warmup followed by cosine decay. All the training was running on an RTX4090 24 GB, using gradient accumulation equivalents.

To enhance generalization, we employ Mixup~\cite{zhang2017mixup} ($\alpha=1.0$), CutMix~\cite{yun2019cutmix} ($\alpha=0.8$), Random Erasing~\cite{zhong2020random} ($p=0.25, ratio=0.2$), Stochastic Depth~\cite{huang2016deep} ($p=0.5$), and Label Smoothing~\cite{muller2019does} ($0.1$). 

\textbf{Datasets.} We evaluate on public audio benchmarks: \textbf{AudioSet} (AS2M/AS20K)~\cite{Audioset}, with $\sim$10\,s clips and mean Average Precision (mAP) as the metric; \textbf{ESC-50} (ESC)~\cite{ESC50}, 5\,s clips evaluated by accuracy; \textbf{Speech Commands v2} (SC~V2)~\cite{SCV2}, $\sim$1\,s clips with 35 classes, evaluated by accuracy; \textbf{NSynth Pitch} (NP)~\cite{nsp}, $\sim$4\,s musical-note clips (50\,h full set), evaluated by accuracy; and \textbf{VGGSound} (VGG)~\cite{vgg}, 10\,s audio-

\textbf{Baselines.} We compare A-RWKV with two sequence modeling operators: (i) Audio Spectrogram Transformer (AST)~\cite{gong2021ast}, representing global self-attention with $\mathcal{O}(L^2)$ cost; and (ii) Audio Mamba (AuM)~\cite{erol2024audio}, a linear-time state-space model. Evaluations use the same training recipe and cover from-scratch training, downstream fine-tuning, scaling, and long-context inference efficiency (latency/throughput).

\subsection{Compare with Other Sequence Modeling Operators.}

\textbf{From-scratch Training.}
As shown in Tab.~\ref{tab:scratch}, A-RWKV-B/16 trained from scratch outperforms both attention-based (AST)~\cite{gong2021ast} and state-space (AuM)~\cite{erol2024audio} baselines on every benchmark reported, indicating consistent generalization across data scales (2K$\sim$2M) and audio domains. AST does not contain convolutional structures and is not data-friendly. Two sets of experiments on AS20K and AS2M intuitively demonstrated this point. Although the AuM also includes a set of Casual Conv1D, this set of Conv1D is not consistent with the original spatial arrangement of the patch, and multiple sets need to be designed in parallel and precisely balanced.

These results highlight the superiority of A-RWKV’s sequence modeling operator over quadratic-time attention and prior linear-time SSMs in the from-scratch setting. Collectively, they suggest that A-RWKV’s operator reconciles global and local structure: Bi-WKV provides full-sequence conditioning with linear complexity, benefiting event-rich, long-context corpora (e.g., AudioSet, VGGSound) while avoiding attention’s quadratic cost and the optimization fragility of prior SSMs, whereas the 2D ConvShift imparts a local time–frequency inductive bias suited to short, discriminative patterns (e.g., Speech Commands V2, NSynth Pitch).

\begin{table}[h!]
\centering
\caption{Results of from-scratch training of AST and AuM base models across various datasets. '*' means our impl. with bf16. All other results are from AuM~\cite{erol2024audio}.}
\resizebox{\linewidth}{!}{
\begin{tabular}{lcccccc}
\toprule
\multirow{2}{*}{\textbf{Model}} & \textbf{AS2M} & \textbf{AS20K} & \textbf{VGG} & \textbf{NP} & \textbf{SC V2} & \textbf{ESC}\\
& \textbf{(mAP)} & \textbf{(mAP)} & \textbf{(Acc.)} & \textbf{(Acc.)} & \textbf{(Acc.)} & \textbf{(Acc.)}\\
\midrule
AST-B/16 & 29.10 & 10.41 & 37.25 & - & 85.27 & -\\
AST-B/16* & 35.23 & 14.25 & 39.88 & 86.31 & 87.31 & 74.2\\
AuM-B/16 & 32.43 & 13.28 & 42.58 & - & 91.58 & -\\
A-RWKV-B/16* & \textbf{40.91} & \textbf{17.25} & \textbf{45.37} & \textbf{91.35} & \textbf{93.01} & \textbf{80.4}\\
\bottomrule
\end{tabular}
}
\label{tab:scratch}
\end{table}

\textbf{Fine-tuning on Downstream Tasks.}
All models are fine-tuned after AudioSet-2M pre-training; the parentheses in Tab.~\ref{tab:finetuning} denote gains over each model’s own from-scratch counterpart. When fed with AS2M pre-training, AST sometimes posts larger deltas, yet its final accuracies remain consistently below A-RWKV-B/16. By contrast, AuM exhibits weaker scaling with data than A-RWKV: on shared benchmarks, A-RWKV not only reaches higher end accuracy but also converts the same pre-training signal into equal-or-larger gains. Notably, although models are pre-trained on AudioSet2M ($\sim$10s clips), A-RWKV delivers greater improvements than AuM when migrated to short-duration datasets such as Speech Commands v2 ($\sim$1s), suggesting better robustness to distributional changes in sequence length.

Beyond the surface comparison of final accuracies, two patterns emerge. First, A-RWKV-B/16 attains the best end performance across all reported benchmarks. Second, while AST can exhibit larger incremental gains, A-RWKV starts from a stronger from-scratch baseline and translates pre-training into consistent—though not always maximal—improvements. A plausible explanation is that A-RWKV couples spectro-temporal locality (2D ConvShift) with linear-time bidirectional context integration (Bi-WKV) atop a stable recurrent backbone, yielding features that fine-tune more reliably.

\begin{table}[h]
\centering
\caption{Fine-tuning performance on downstream tasks. All models are pre-trained on AudioSet-2M. Accuracies are reported in percent (\%). '*' means our impl. with bf16.}
\label{tab:finetuning}
\resizebox{\columnwidth}{!}{%
\begin{tabular}{lcccc}
\toprule
\textbf{Model} & \textbf{NP} & \textbf{SC V2} & \textbf{ESC} & \textbf{VGG} \\
\midrule
AST-B/16 & 90.15(+3.84)* & 90.37(+5.10) & 83.5(+9.3)* & 44.17(+6.92) \\
AuM-B/16 & - & 94.78(+3.20) & - & 46.61(+4.03) \\
A-RWKV-B/16 & 93.44(+2.07) & 96.83(+3.82) & 86.8(+6.2) & 48.91(+3.54) \\
\bottomrule
\end{tabular}%
}
\end{table}

\textbf{Scaling Analysis.} For completeness, the lightweight variants use modest regularization and dimensions: \textbf{A-RWKV-T} employs stochastic depth $p{=}0.05$, cutmix $\alpha=0.2$, and embedding dimension $192$; \textbf{A-RWKV-S} employs stochastic depth $p{=}0.35$, cutmix $\alpha=1.0$, and embedding dimension $384$. As summarized in Tab.~\ref{tab:finetuning}, A-RWKV scales smoothly from tiny to small to base with consistent gains across benchmarks, aligning with our motivation for a stable, linear-time sequence operator. 

\begin{table}[h!]
\centering
\caption{Scaling and downstream fine-tuning with A-RWKV. All models are pre-trained on AudioSet-2M. Metrics are mAP for AudioSet-2M and accuracy (\%) for VGGSound and ESC-50.}
\label{tab:scaling}
\resizebox{\columnwidth}{!}{%
\begin{tabular}{l|c|ccc}
\toprule
\textbf{Model} & \textbf{Params} & \textbf{AudioSet 2M} & \textbf{VGGSound} & \textbf{ESC-50} \\
\midrule
AST-B & 86M & 35.23 & 39.88 & 74.2 \\
AuM-B & 92M & 32.43 & 42.58 & - \\
\midrule
\textbf{A-RWKV-T (Ours)} & \textbf{6M} & \textbf{30.07} & \textbf{39.82} & \textbf{74.6} \\
\textbf{A-RWKV-S (Ours)} & \textbf{23M} & \textbf{37.84} & \textbf{43.15} & \textbf{77.5} \\
\textbf{A-RWKV-B (Ours)} & \textbf{91M} & \textbf{40.91} & \textbf{45.37} & \textbf{80.4} \\
\bottomrule
\end{tabular}%
}
\end{table}
Performance varies by dataset: the -T variant lags on AudioSet likely because multi-class classification needs more diverse features than its limited hidden dimension can capture. Meanwhile, A-RWKV stays competitive even at tiny scale—its small model broadly surpasses attention baselines, and the base model widens the lead and beats state-space counterparts—showing that the Bi-WKV operator, with RWKV-style stability, scales reliably and models long contexts without the training fragility seen in large SSMs.

\subsection{Model Efficiency and Ablation.}

\begin{figure}[h!]
    \centering
    \subfigure[GPU Inference Runtime]{
        \centering
        \includegraphics[width=\linewidth]{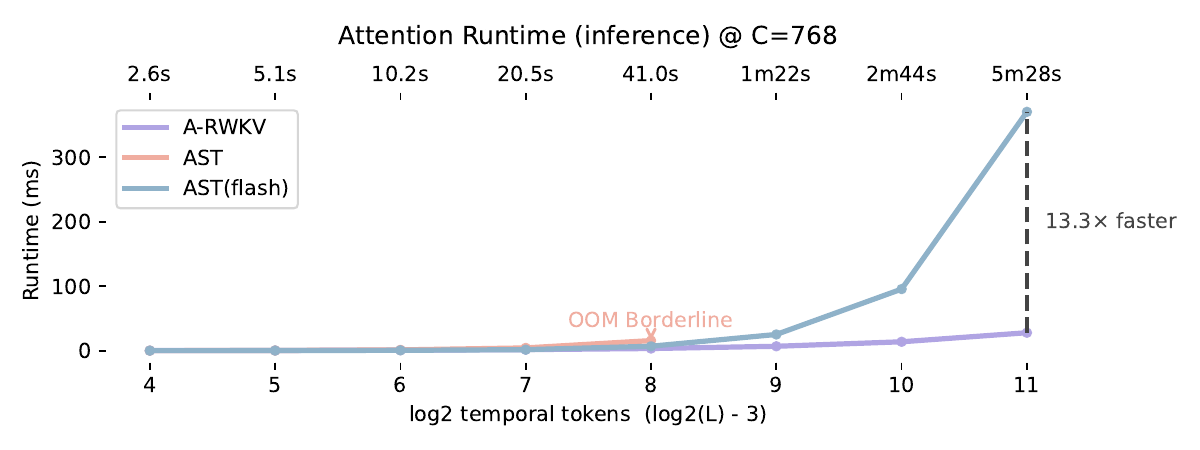}
        \label{fig:gpu_speed}
    }
    \hfill
    \subfigure[Token Per Second]{
        \centering
        \includegraphics[width=\linewidth]{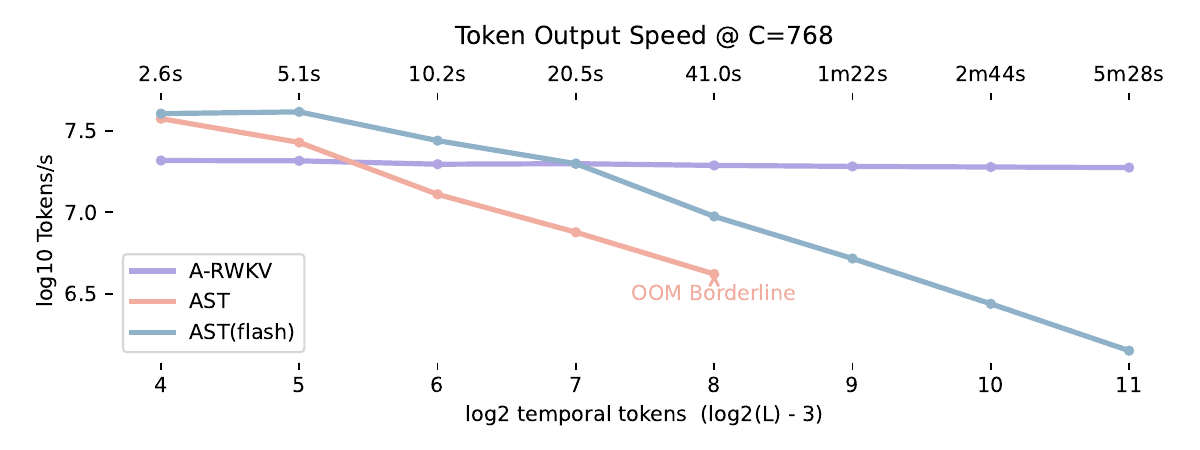}
        \label{fig:token_speed}
    }
    \vfill
    \caption{Efficiency analysis of A-RWKV's RWKV7 operator against baselines. We compare (a) GPU inference speed, and (b) token per second as input audio sequence length increases.}
    \label{fig:efficiency}
\end{figure}

\textbf{Efficiency Analysis.}
As shown in Fig.~\ref{fig:efficiency}, with $C=768$ and sequence length from $2^4$ to $2^{11}$ temporal tokens ($\sim$2.6s to $\sim$5m28s audio), A-RWKV's operator (RWKV7) matches AST/AST(flash) at short context but scales much more gently as tokens grow. AST shows OOM near $2^8$ ($\sim$20.5 s), and AST(flash)~\cite{dao2022flashattention} latency rises steeply; at the longest length A-RWKV is $\sim 13.3\times$ faster than AST(flash). 
In token throughput (log10), A-RWKV remains almost flat while both AST variants degrade and approach OOM at mid–long contexts. Despite its kernel-level improvements, FlashAttention’s efficiency at large scales remains fundamentally bounded by the required number of attention computations. 

Under lower tokens, since the RNN-Style model is pseudo-parallel, the difference is not significant compared with highly parallelized attention computing.
Prior work~\cite{erol2024audio, zhu2024vision} has already reported extensive operator-level comparisons. Overall, A-RWKV delivers stable compute/memory behavior and clearly superior long-context efficiency.

\textbf{Ablation Study.} As shown in Tab.~\ref{tab:ablation}, replacing the causal RWKV7 operator with bidirectional scanning (Bi-WKV) yields the most salient improvement and aligns with our original motivation. RNN-style models have an inherent weakness: they tend to forget earlier information, whereas labels in audio may emerge at arbitrary moments; cues that surface early in long spectrograms are therefore easily discounted~\cite{sieber2024understanding}. Introducing a backward scan restores these early signals and enables a more faithful global summary; for a similar reason, some designs~\cite{zhu2024vision} even place the [CLS] token near the middle to balance information flow rather than at an endpoint. A lightweight fusion gate further strengthens the bidirectional design by adaptively weighting forward and backward evidence, effectively emphasizing whichever temporal direction better explains the current segment, especially when decisive cues occur away from sequence boundaries.

\begin{table}[h!]
\centering
\caption{Ablation study on the key components of AudioRWKV on the AS2M dataset. "Original" token shift refers to the 1D method in RWKV7. (F) is full AudioRWKV.}
\label{tab:ablation}
\resizebox{\columnwidth}{!}{%
\begin{tabular}{l|cccc|c}
\toprule
\textbf{Variant} & \textbf{Scanning} & \textbf{Fusion Gate} & \textbf{Token Shift} & \textbf{Aug.} & \textbf{mAP.} \\
\midrule
(A) RWKV7 & Causal & - & Original & - & 34.50 \\
(B) + Cutmix & Causal & - & Original & \checkmark & 34.89 \\
(C) + Bi-Scan & Bi & Average & Original & \checkmark & 38.39 \\
(D) + Gate & Bi & Weighted & Original & \checkmark & 39.02 \\
(E) + Q-Shift & Bi & Weighted & Q-Shift & \checkmark & 39.35 \\
(F) + Conv & Bi & Weighted & ConvShift & \checkmark & 40.91 \\
\bottomrule
\end{tabular}%
}
\end{table}

Beyond directionality, we move from the original one-dimensional token shift to a two-dimensional depthwise-separable ConvShift, instantiated via a simpler Q-Shift variant inspired by V-RWKV~\cite{duan2024vision}. Conceptually, the convolution behaves like an adaptive token shift over local time–frequency neighborhoods, aligning spectro-temporal patterns while preserving the linear computational profile and the training stability characteristic of RWKV7. The full A-RWKV variant unifies stable recurrence, linear-time global context modeling, and locality-aware two-dimensional structure, and consequently surpasses the causal baseline across diverse audio regimes.

\section{Conclusion}
Audio-RWKV performs as well as standard Transformers and Mambas in audio tasks, while using much less computation and memory, demonstrating the promise of linear models for audio understanding. Although inductive biases and explicit decay are beneficial in low-to-moderate data regimes, they may impair performance at sufficiently large scales~\cite{tolstikhin2021mlp, zhai2022scaling}, whose precise threshold is difficult to determine, and they are ill-suited for incorporating ViT-based (AST) masked pre-training~\cite{he2022masked}. Further, reducing a 2D spatial field to a 1D sequence destroys locality, such that sequential neighbors need not be spatial neighbors~\cite{xiao2024spatial}; token shift compensates and rethinking scanning sequence~\cite{huang2024localmamba} only partially. We will pursue comprehensive remedies in future work.

\newpage
\bibliographystyle{IEEEbib}
\bibliography{refs}

@article{vaswani2017attention,
  title={Attention is all you need},
  author={Vaswani, Ashish and Shazeer, Noam and Parmar, Niki and Uszkoreit, Jakob and Jones, Llion and Gomez, Aidan N and Kaiser, {\L}ukasz and Polosukhin, Illia},
  journal={Advances in neural information processing systems},
  volume={30},
  year={2017}
}

@article{gong2021ast,
  title={Ast: Audio spectrogram transformer},
  author={Gong, Yuan and Chung, Yu-An and Glass, James},
  journal={arXiv preprint arXiv:2104.01778},
  year={2021}
}

@article{dosovitskiy2020image,
  title={An image is worth 16x16 words: Transformers for image recognition at scale},
  author={Dosovitskiy, Alexey and Beyer, Lucas and Kolesnikov, Alexander and Weissenborn, Dirk and Zhai, Xiaohua and Unterthiner, Thomas and Dehghani, Mostafa and Minderer, Matthias and Heigold, Georg and Gelly, Sylvain and others},
  journal={arXiv preprint arXiv:2010.11929},
  year={2020}
}

@article{gu2023mamba,
  title={Mamba: Linear-time sequence modeling with selective state spaces},
  author={Gu, Albert and Dao, Tri},
  journal={arXiv preprint arXiv:2312.00752},
  year={2023}
}

@article{erol2024audio,
  title={Audio mamba: Bidirectional state space model for audio representation learning},
  author={Erol, Mehmet Hamza and Senocak, Arda and Feng, Jiu and Chung, Joon Son},
  journal={IEEE Signal Processing Letters},
  year={2024},
  publisher={IEEE}
}

@article{zhu2024vision,
  title={Vision mamba: Efficient visual representation learning with bidirectional state space model},
  author={Zhu, Lianghui and Liao, Bencheng and Zhang, Qian and Wang, Xinlong and Liu, Wenyu and Wang, Xinggang},
  journal={arXiv preprint arXiv:2401.09417},
  year={2024}
}

@article{duan2024vision,
  title={Vision-rwkv: Efficient and scalable visual perception with rwkv-like architectures},
  author={Duan, Yuchen and Wang, Weiyun and Chen, Zhe and Zhu, Xizhou and Lu, Lewei and Lu, Tong and Qiao, Yu and Li, Hongsheng and Dai, Jifeng and Wang, Wenhai},
  journal={arXiv preprint arXiv:2403.02308},
  year={2024}
}

@article{peng2023rwkv,
  title={Rwkv: Reinventing rnns for the transformer era},
  author={Peng, Bo and Alcaide, Eric and Anthony, Quentin and Albalak, Alon and Arcadinho, Samuel and Biderman, Stella and Cao, Huanqi and Cheng, Xin and Chung, Michael and Grella, Matteo and others},
  journal={arXiv preprint arXiv:2305.13048},
  year={2023}
}

@article{sun2023retentive,
  title={Retentive network: A successor to transformer for large language models},
  author={Sun, Yutao and Dong, Li and Huang, Shaohan and Ma, Shuming and Xia, Yuqing and Xue, Jilong and Wang, Jianyong and Wei, Furu},
  journal={arXiv preprint arXiv:2307.08621},
  year={2023}
}

@article{peng2025rwkv,
  title={Rwkv-7" goose" with expressive dynamic state evolution},
  author={Peng, Bo and Zhang, Ruichong and Goldstein, Daniel and Alcaide, Eric and Du, Xingjian and Hou, Haowen and Lin, Jiaju and Liu, Jiaxing and Lu, Janna and Merrill, William and others},
  journal={arXiv preprint arXiv:2503.14456},
  year={2025}
}

@inproceedings{li2024exploring,
  title={Exploring Receptance Weighted Key Value Model for Single-Channel Speech Enhancement},
  author={Li, Yuanle and Zhou, Yi and Liu, Hongqing},
  booktitle={2024 7th International Conference on Information Communication and Signal Processing (ICICSP)},
  pages={123--127},
  year={2024},
  organization={IEEE}
}

@article{xiao2025rwkvtts,
  title={RWKVTTS: Yet another TTS based on RWKV-7},
  author={Xiao, Liu and others},
  journal={arXiv preprint arXiv:2504.03289},
  year={2025}
}

@article{sieber2024understanding,
  title={Understanding the differences in foundation models: Attention, state space models, and recurrent neural networks},
  author={Sieber, Jerome and Alonso, Carmen A and Didier, Alexandre and Zeilinger, Melanie N and Orvieto, Antonio},
  journal={Advances in Neural Information Processing Systems},
  volume={37},
  pages={134534--134566},
  year={2024}
}

@INPROCEEDINGS{Audioset,
  author={Gemmeke, Jort F. and Ellis, Daniel P. W. and Freedman, Dylan and Jansen, Aren and Lawrence, Wade and Moore, R. Channing and Plakal, Manoj and Ritter, Marvin},
  booktitle={IEEE International Conference on Acoustics, Speech and Signal Processing (ICASSP)}, 
  title={Audio Set: An ontology and human-labeled dataset for audio events}, 
  year={2017},
  volume={},
  number={},
  pages={776-780}
}

@article{SCV2,
  title={Speech Commands: A Dataset for Limited-Vocabulary Speech Recognition},
  author={Pete Warden},
  journal={ArXiv},
  year={2018},
  volume={abs/1804.03209}
}

@inproceedings{ESC50,
  title={ESC: Dataset for Environmental Sound Classification},
  author={Karol J. Piczak},
  booktitle={ACM international conference on Multimedia (ACM MM)},
  year={2015}
}

@inproceedings{nsp,
  title={Neural audio synthesis of musical notes with wavenet autoencoders},
  author={Engel, Jesse and Resnick, Cinjon and Roberts, Adam and Dieleman, Sander and Norouzi, Mohammad and Eck, Douglas and Simonyan, Karen},
  booktitle={International conference on machine learning (ICML)},
  pages={1068--1077},
  year={2017},
  organization={PMLR}
}

@inproceedings{vgg,
  title={Vggsound: A large-scale audio-visual dataset},
  author={Chen, Honglie and Xie, Weidi and Vedaldi, Andrea and Zisserman, Andrew},
  booktitle={ICASSP 2020-2020 IEEE International Conference on Acoustics, Speech and Signal Processing (ICASSP)},
  pages={721--725},
  year={2020},
  organization={IEEE}
}

@article{zhang2017mixup,
  title={mixup: Beyond empirical risk minimization},
  author={Zhang, Hongyi and Cisse, Moustapha and Dauphin, Yann N and Lopez-Paz, David},
  journal={arXiv preprint arXiv:1710.09412},
  year={2017}
}

@inproceedings{yun2019cutmix,
  title={Cutmix: Regularization strategy to train strong classifiers with localizable features},
  author={Yun, Sangdoo and Han, Dongyoon and Oh, Seong Joon and Chun, Sanghyuk and Choe, Junsuk and Yoo, Youngjoon},
  booktitle={Proceedings of the IEEE/CVF international conference on computer vision},
  pages={6023--6032},
  year={2019}
}

@article{muller2019does,
  title={When does label smoothing help?},
  author={M{\"u}ller, Rafael and Kornblith, Simon and Hinton, Geoffrey E},
  journal={Advances in neural information processing systems},
  volume={32},
  year={2019}
}

@inproceedings{zhong2020random,
  title={Random erasing data augmentation},
  author={Zhong, Zhun and Zheng, Liang and Kang, Guoliang and Li, Shaozi and Yang, Yi},
  booktitle={Proceedings of the AAAI conference on artificial intelligence},
  volume={34},
  number={07},
  pages={13001--13008},
  year={2020}
}

@inproceedings{adamw,
  title={Decoupled Weight Decay Regularization},
  author={Loshchilov, Ilya and Hutter, Frank},
  booktitle={International Conference on Learning Representations (ICLR)},
  year={2019},
}

@inproceedings{huang2016deep,
  title={Deep networks with stochastic depth},
  author={Huang, Gao and Sun, Yu and Liu, Zhuang and Sedra, Daniel and Weinberger, Kilian Q},
  booktitle={European conference on computer vision},
  pages={646--661},
  year={2016},
  organization={Springer}
}

@article{dao2022flashattention,
  title={Flashattention: Fast and memory-efficient exact attention with io-awareness},
  author={Dao, Tri and Fu, Dan and Ermon, Stefano and Rudra, Atri and R{\'e}, Christopher},
  journal={Advances in neural information processing systems},
  volume={35},
  pages={16344--16359},
  year={2022}
}

@inproceedings{zhai2022scaling,
  title={Scaling vision transformers},
  author={Zhai, Xiaohua and Kolesnikov, Alexander and Houlsby, Neil and Beyer, Lucas},
  booktitle={Proceedings of the IEEE/CVF conference on computer vision and pattern recognition},
  pages={12104--12113},
  year={2022}
}

@article{tolstikhin2021mlp,
  title={Mlp-mixer: An all-mlp architecture for vision},
  author={Tolstikhin, Ilya O and Houlsby, Neil and Kolesnikov, Alexander and Beyer, Lucas and Zhai, Xiaohua and Unterthiner, Thomas and Yung, Jessica and Steiner, Andreas and Keysers, Daniel and Uszkoreit, Jakob and others},
  journal={Advances in neural information processing systems},
  volume={34},
  pages={24261--24272},
  year={2021}
}

@article{xiao2024spatial,
  title={Spatial-mamba: Effective visual state space models via structure-aware state fusion},
  author={Xiao, Chaodong and Li, Minghan and Zhang, Zhengqiang and Meng, Deyu and Zhang, Lei},
  journal={arXiv preprint arXiv:2410.15091},
  year={2024}
}

@inproceedings{huang2024localmamba,
  title={Localmamba: Visual state space model with windowed selective scan},
  author={Huang, Tao and Pei, Xiaohuan and You, Shan and Wang, Fei and Qian, Chen and Xu, Chang},
  booktitle={European Conference on Computer Vision},
  pages={12--22},
  year={2024},
  organization={Springer}
}

@inproceedings{he2022masked,
  title={Masked autoencoders are scalable vision learners},
  author={He, Kaiming and Chen, Xinlei and Xie, Saining and Li, Yanghao and Doll{\'a}r, Piotr and Girshick, Ross},
  booktitle={Proceedings of the IEEE/CVF conference on computer vision and pattern recognition},
  pages={16000--16009},
  year={2022}
}

\end{document}